%% file: main.tex
\documentclass[fleqn,usenatbib]{mnras}

\usepackage{newtxtext,newtxmath}
\usepackage[T1]{fontenc}
\usepackage{pdflscape}
\usepackage{longtable}
\usepackage{booktabs}
\usepackage{color}
\usepackage{mathtools}
\usepackage{xspace}
\usepackage{rotating}
\usepackage{appendix}
\usepackage{multirow}
\usepackage{multicol}
\usepackage{array}
\usepackage{subcaption}
\usepackage{comment}
\usepackage{tabularx}
\usepackage{bigstrut}
\usepackage{threeparttable}
\usepackage{afterpage}
\usepackage{capt-of}
\usepackage{newtxtext,newtxmath}
\usepackage[T1]{fontenc}
\usepackage{ae,aecompl}
\usepackage{graphicx}	
\usepackage{amsmath}	
\usepackage{blindtext}
\usepackage{comment}
\usepackage{tabularx}
\usepackage{tabulary}
\usepackage{bigstrut}
\usepackage{threeparttable}
\usepackage{afterpage}
\usepackage{capt-of}
\usepackage{caption}
\usepackage{subcaption}
\usepackage{hyperref}

\DeclareRobustCommand{\VAN}[3]{#2}
\let\VANthebibliography\thebibliography
\def\thebibliography{\DeclareRobustCommand{\VAN}[3]{##3}\VANthebibliography}

   \newcommand{\ecyc}{E$_{\texttt{cyc}}$} 
   \newcommand{\Lx}{$L_{\texttt{X}}$}
   \newcommand{\Lc}{$L_{\texttt{crit}}$}
   \newcommand{\funi}{erg cm$^{-2}$ s$^{-1}$}
\usepackage{graphicx}	% Including figure files
\usepackage{amsmath}	% Advanced maths commands

\title[\textit Discovery of a cyclotron line in GRO J1750-27]{\textit{NuSTAR} discovery of a cyclotron line in GRO J1750-27}
\author[Devaraj \& Paul]{Ashwin Devaraj$^{1,2}$\thanks{\href{mailto:ashwin@rri.res.in}{ashwin@rri.res.in}} and Biswajit Paul$^{1}$\thanks{ \href{mailto:bpaul@rri.res.in}
{bpaul@rri.res.in}}\\
$^{1}$Raman Research Institute, Sadashivnagar, Bangalore-560080, India\\
$^{2}$Joint Astronomy Programme, Indian Institute of Science, Bangalore-560012, India\\}

\date{Accepted XXX. Received YYY; in original form ZZZ}

\pubyear{2015}

\begin{document}
\label{firstpage}
\pagerange{\pageref{firstpage}--\pageref{lastpage}}
\maketitle

\begin{abstract}
GRO J1750-27, discovered during an outburst in 1995 with CGRO-BATSE, is one of the farthest known galactic Be-X-ray binary systems. This relatively poorly studied system recently went into an outburst in September 2021. The source was observed during the latest outburst using the \textit{NuSTAR} telescope during the rising phase of the outburst. We estimate the spin period of the source to be 4.45 s using which we produced energy-resolved pulse profiles between 3 and 65 keV. We find that the profile is double-peaked at low energies (<18 keV) while evolving into a single peak at higher energies (>18 keV). The broadband spectrum of this source was fitted well with a high energy cutoff power-law model and we report the discovery of a cyclotron resonant scattering feature (CRSF) in this source at 43 keV indicating a magnetic field strength of $3.7 \times 10^{12}$ G. Our estimate of the magnetic field strength using the cyclotron line is consistent with the estimates made earlier using the accretion torque model from measurements of spin-up rates and fluxes during the previous outbursts.

\end{abstract}

% Select between one and six entries from the list of approved keywords.AX J1749.1−2639

%%%%%%%%%%%%%%%%%%%%%%%%%%%%%%%%%%%%%%%%%%%%%%%%%%

%%%%%%%%%%%%%%%%% BODY OF PAPER %%%%%%%%%%%%%%%%%%

\section{INTRODUCTION}
Be/X-ray binaries  (BeXRB), a sub-class of High Mass X-ray Binaries  (HMXB) are systems that host a Neutron Star  (NS) and a fast rotating B-type main sequence companion that also exhibit emission lines from the circumstellar disc around the companion  \citep{Reig_2011}. GRO J1750-27  (also known as AX J1749.1-2639) is one such transient system discovered with the CGRO/BATSE during an outburst in 1995, that lasted about two months \citep{Koh_1995,Scott_1997}. Coherent pulsations of 4.45 s interpreted as the spin period and a 29.8 day orbital period were reported by  \citet{Scott_1997} using the same data. Though no optical counterpart was detected, based on the position of this source on the Corbet diagram  \citep{corbet_1984},  \citet{Scott_1997} concluded that the system must be a BeXRB. 
\par
The source once again underwent an outburst around Jan. 2008, which lasted for $\sim$ 150 days, during which it was observed with \textit{INTEGRAL} and \textit{Swift XRT}   \citep{Shaw_2009}. Along with providing an updated orbital ephemeris,  \citet{Shaw_2009} also modelled the broadband spectrum of GRO J1750-27 with a cutoff power-law, estimating a flux of 6.5$\times$10$^{-9}$ \funi in the 0.1 - 100 keV range. Using the spin-up rate and flux determined from the  \textit{Swift/BAT} data and applying the Ghosh \& Lamb model  \citep{Ghosh_1978}, they estimated a surface magnetic field of B $\sim 2\times 10^{12}$ G and a distance range of 12 - 22 kpc.
\par 
This source's third major outburst began in Dec. 2014 and lasted up to May 2015  \citep{Finger_2014,Lutovinov_2019}.  During this period it was also observed with \textit{Chandra, Swift} and \textit{Fermi} from which  \citet{Lutovinov_2019} determined a source position of $\text { R.A. }=17^{\mathrm{h}} 49^{\mathrm{m}} 12.99^{\mathrm{s}}, \text { Dec. }=-26^{\circ} 38^{\prime} 38.5^{\prime \prime}$  (in the J2000 system) and identified an IR counterpart at a distance of >12 kpc. Analysing the spin evolution of GRO J1750-27, they also provided an independent distance estimate between 14 and 22 kpc with the expected magnetic field strength to be in the range  (3.5 - 4.5)$\times 10^{12}$ G. Using data from the VVV/ESO and UKIDSS/GPS surveys, \citet{Lutovinov_2019} also concluded that the companion must be an early B-type star. Using Swift/XRT and Chandra data  \citet{Escorial_2019} investigated the cooling of the heated NS crust during quiescence after the 2015 outburst and concluded that the X-ray emission is likely due to low-level accretion rather than being due to the cooling.    

 \quad Once again, on 18 September 2021, GRO J1750-27 began to exhibit another major outburst and coherent pulsations of 4.45 s were detected by \textit{Fermi/GBM}  \citep{Malacaria_2021}.  In this work, we investigated this source's spectral and timing characteristics using a \textit{NuSTAR} observation that was made during this outburst and report the discovery of a cyclotron line at $\sim$ 43 keV.  

\begin{figure}
    \centering
	\includegraphics[scale=0.32,trim={0 1.0cm 0 2.0cm},  angle=-90]{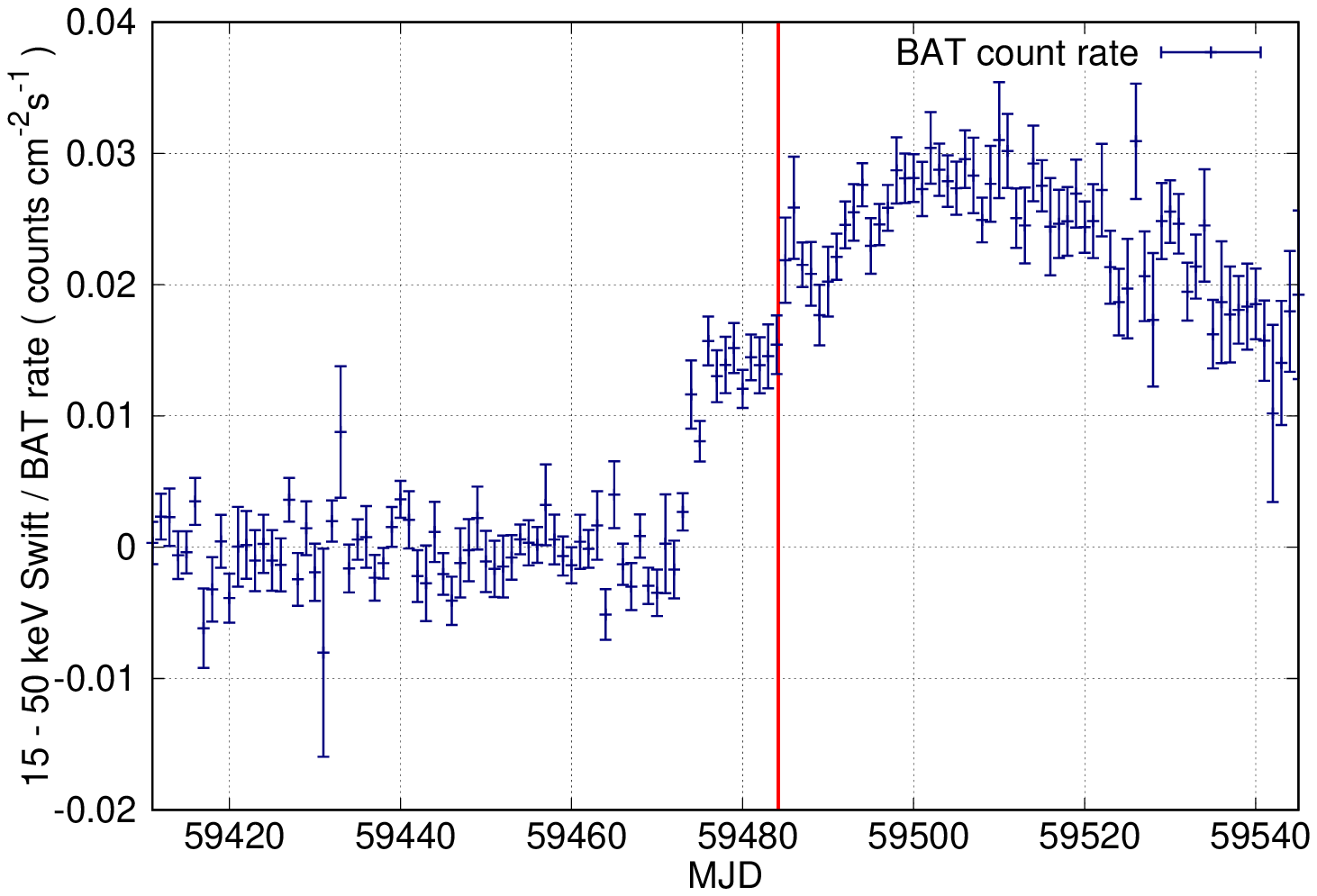}
    \caption{\textit{Swift/BAT} light curve of the 2021 outburst in the 15 - 50 keV range. Blue points indicate the \textit{Swift/BAT} count rate while the vertical red line is indicative of when the \textit{NuSTAR} observation was made. The data were obtained from \url{https://swift.gsfc.nasa.gov/results/transients/weak/AXJ1749.1-2639}}
    \label{fig:outburst-lightcurve} 
\end{figure}

\section{OBSERVATIONS AND DATA REDUCTION}
The most recent outburst of this source began on 18 September 2021, and was still ongoing at the time of the preparation of this manuscript. 15 - 50 keV \textit{Swift/BAT} light curves of the source's outburst are available, and a 30 ks \textit{NuSTAR} observation was made on the 27$^{\mathrm{th}}$ September 2021  (OBSID:90701331002), during the rising phase of the outburst  (see Fig. \ref{fig:outburst-lightcurve}). The source continued to brighten until it reached a maximum flux of almost twice the flux measured during the time of this \textit{NuSTAR} observation. 
\par
\begin{figure*}
    %`\centering
%	\includegraphics[scale=0.85, trim={0 1.0cm 0 2.0cm}]{finalSourceimg.eps}
    \includegraphics[width = \textwidth,height= 7.5cm]{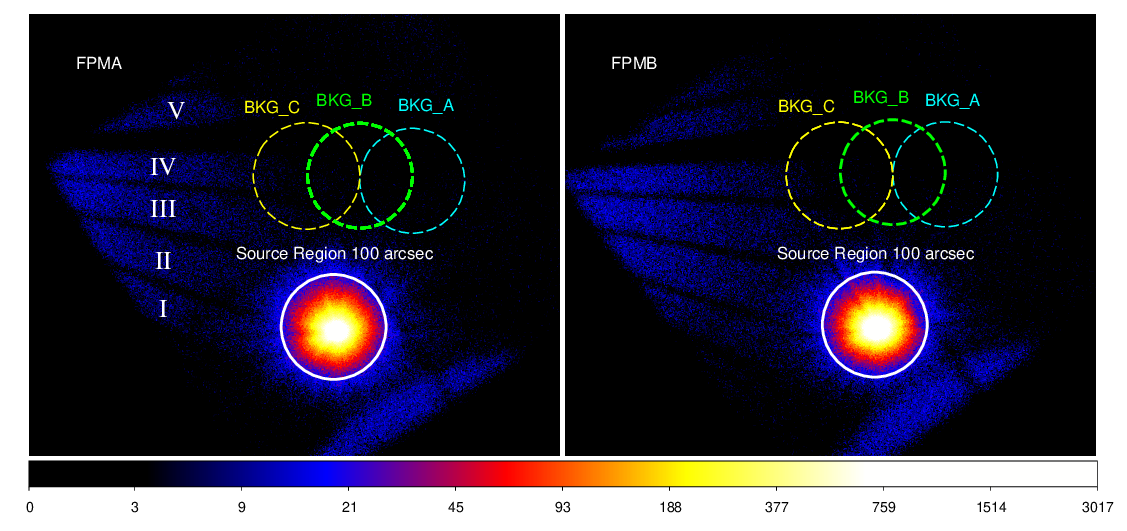}
    \caption{Image of the source obtained with FPMA  (left) and FPMB  (right), scaled logarithmically, are shown along with the source extraction region  (white) and three different choices of background regions BKG\_A  (blue), BKG\_B  (green) and BKG\_C  (yellow) each of 100 arc seconds radius. The five streaks of Ghost Rays seen in both FPMA and FPMB have been named using Roman Numerals.}
    \label{fig:SourceImage} 
\end{figure*}
The \textit{Nuclear Spectroscopic Telescope Array}    (\textit{NuSTAR}) is a hard X-ray telescope with two identical focal plane modules  (FPMA and FPMB), each containing four solid-state CdZnTe detectors, sensitive in the 3 - 79 keV range  \citep{Harrison_2013}. Its spectral resolution is $\sim$ 0.4 keV at 10 keV and $\sim$ 0.9 keV at 60 keV. The X-ray optics is a conical approximation of a Wolter I geometry in which the X-rays are focused onto the detectors through grazing incidence reflection. There are cases where the X-rays from off-axis sources get reflected off only one of the mirrors before reaching the detectors, and these give rise to Ghost rays in the image  \citep{Madsen_2017}. This \textit{NuSTAR} observation of GRO J1750-27 suffers from this effect, as can be seen in Fig. \ref{fig:SourceImage}. 
\par
\begin{figure}
    \centering
	\includegraphics[scale=0.30,trim={0 1.0cm 0 2.0cm},  angle=-90]{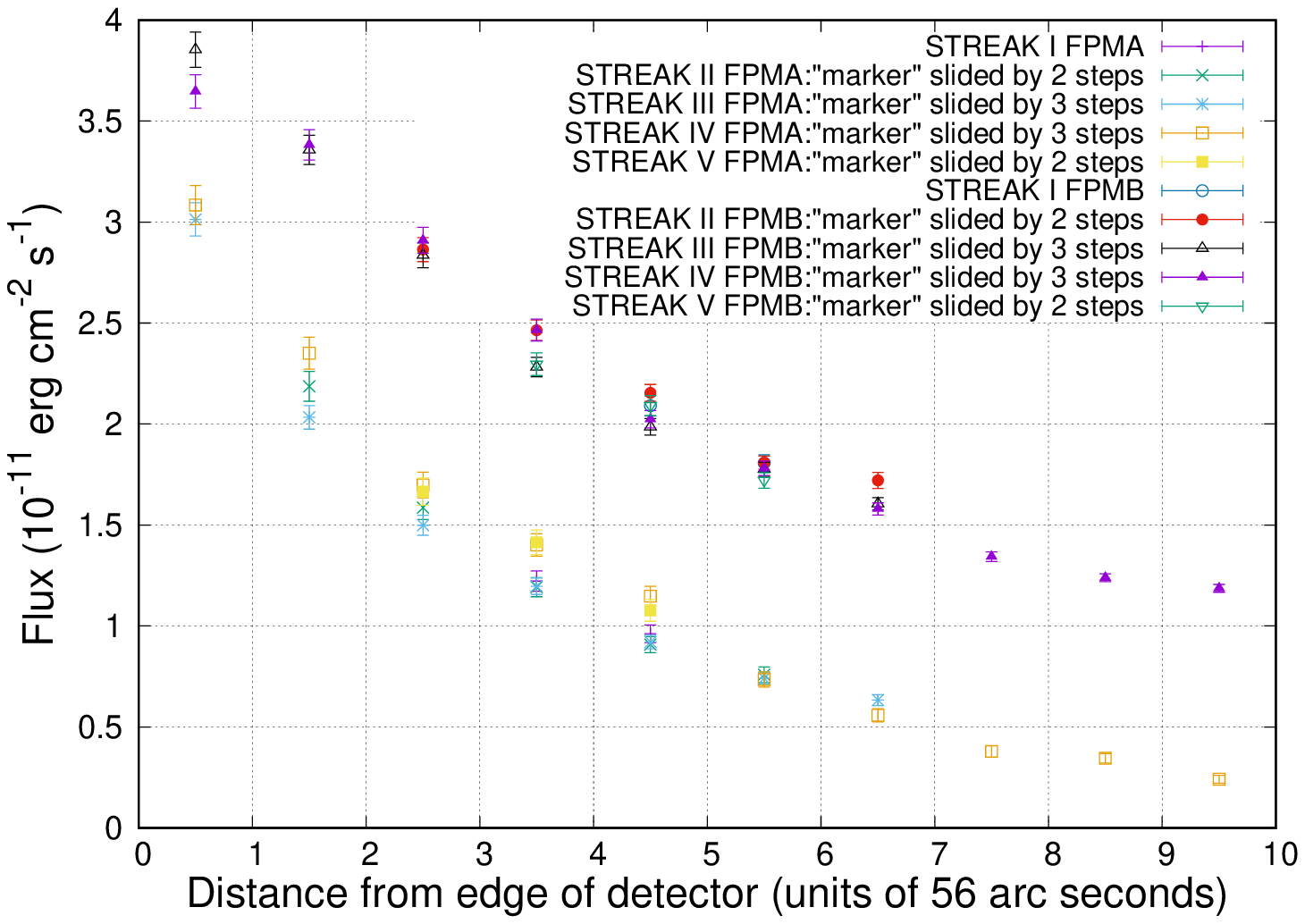}
    \caption{Measurements of X-ray flux along the Ghost Ray streaks in FPMA and FPMB are shown here. The data points for all FPMB regions are plotted after adding a constant number and can be seen shifted above the FPMA data points. Fluxes were measured from consecutive circular regions of 28 arcsec radius, starting from the left edge of each streak. All the measurements along any streak were then slided by one or more steps (of 56 arcsec) to have the flux distribution of all streaks aligned as shown here. The source extraction region is 7 and 9 steps away from the edge of the detector on streak I and streak II respectively (which are not shown in the plot), and the background rate at the source location can therefore be considered to be similar to the region 10 steps away from the edge of the detector along streaks III and IV, which are marked with large circle (BKG\_B) in the Fig. \ref{fig:SourceImage} }
    \label{fig:Flux-streaks} 
\end{figure}
\begin{figure}
    \centering
	\includegraphics[scale=0.335,trim={0 1.2cm 0 2.0cm},  angle=-90]{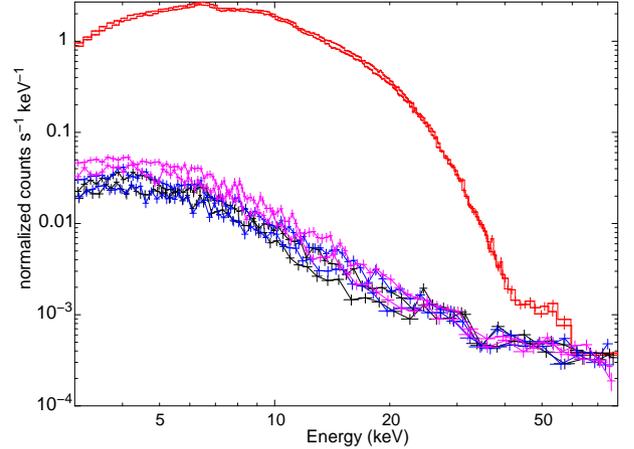}
	%\centering
    \caption{The red curves represent spectrum of the white source region of both FPMA and FPMB  without the subtraction of the background.Similarly, the Black, Blue and Magenta curves represent the spectra from the regions BKG\_A, BKG\_B and BKG\_C respectively  (see Fig \ref{fig:SourceImage})}
    \label{fig:SourceBkgSpectrum} 
\end{figure}

The \textit{NuSTAR} data was reduced in the standard way using \textsc{heasoft} v6.29c, \textsc{nustardas} v2.1.1   (CALDB version:20211020). The \texttt{nupipeline} script was used to produce the filtered event files. The source region was selected with a radius of 100 arc seconds as shown in Fig. \ref{fig:SourceImage}. Due to the source region being contaminated by the photons from the Ghost Rays  (GRs), the choice of background needed to be made
carefully, and from Fig. \ref{fig:SourceImage} it can be seen that the GRs are similarly present in both FPMA and FPMB. Streak \texttt{II} and a portion of Streak \texttt{I} contaminate the source region. From Fig. \ref{fig:SourceImage} it is clear that the intensity drops as a function of the distance from the edge of the detector. Selecting consecutive circular regions of 28 arc seconds radius along each of the streaks, we estimated the flux in each of the regions. We found that the drop in the flux along each of the streaks followed a similar trend (see Fig. \ref{fig:Flux-streaks}). Based on this we selected a region of 100 arc second radius  (BKG\_B) along Streaks \texttt{III} and \texttt{IV} such that the flux there should be approximately the amount by which Streaks \texttt{I} and \texttt{II} contaminate the source region. Two more regions adjacent to BKG\_B, i.e., BKG\_A and BKG\_C, were selected along the Streaks \texttt{III} and \texttt{IV} to understand how sensitive the spectral analysis would be to the choice of background region. The spectra of the source and three background regions were extracted  using the \texttt{nuproducts} script and can be seen in Fig. \ref{fig:SourceBkgSpectrum}. It is clear from the figure that the black and blue curves that represent the spectra from regions BKG\_A and BKG\_B are very similar while the magenta curves representing the spectra from region BKG\_C is slightly higher than the other two up to $\sim$ 20 keV. Above 20 keV, all the backgrounds are quite similar. We limit the spectral analysis to the 3 - 65 keV range as the source is not detectable above 65 keV.  
\par
For the timing analysis, the \texttt{nuproducts} script was used on the barycenter corrected, filtered event files to produce the light curves with a time resolution of 0.01s for both FPMA and FPMB. For the spectral analysis, the spectra were optimally binned based on the scheme given by  \citet{Kaastra_2016} and fit simultaneously with the relative normalisation between the two modules being allowed to freely vary  (fixing the FPMA's normalisation constant to unity). This relative  normalisation was found to be C$_B \sim 1.01$. The spectral analysis was carried out using \textit{XSPEC} version 12.12.0.

\section{ANALYSIS AND RESULTS}
\subsection{Timing Analysis}

We found the spin period of the source, P$_{\mathrm{spin}}$ = 4.451271 (2) s using the epoch folding $\chi^2$ maximization technique using the \texttt{efsearch} tool which is a part of the FTOOLS package.

%%%%%%%%%%%%%%%%%%%%%%%%%%%%%____ENERGY RESOLVED PULSE PROFILES_______________%%%%%%%%%%%%%%%%%%%%%%%%%%%%5

\begin{figure}
\centering
\vspace{0.6 cm}
\includegraphics[height=8cm,width=8cm,trim={1.18cm 2.47cm 1.5cm 0.7cm},clip ]{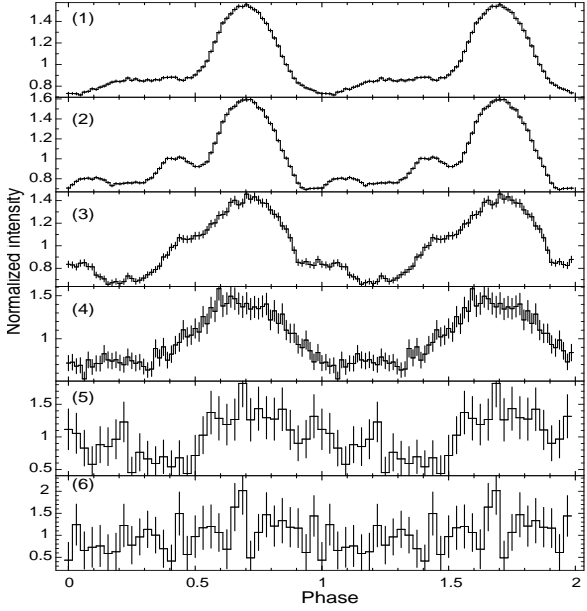} \caption{Energy resolved pulse profiles. The six panels represent the pulse profiles in the respective energy ranges:   (1) 3 - 9  keV,   (2) 9 - 18  keV,  (3) 18 - 28  keV,  (4) 28 - 38  keV,  (5) 38 - 48  keV,  (6) 48 - 65  keV }
\label{fig:BBpulseprofiles-all}
%\vspace{0.1cm}
\hspace{5mm}
\end{figure}

%%%%%%%%%%%%%%%%%%%%%%%%%%%%%%%%%%%%%%%%%%%%%%%%%%%%%%%%%%%%%%%%%%%%%%%%%%%%%%%%%%%%%%%%%%%%%%%%%%%%%%%%%%%%5

\subsubsection{Energy-resolved Pulse Profiles}

The energy-resolved pulse profiles were generated for the 3 - 9  keV,  9 - 18  keV, 18 - 28  keV, 28 - 38  keV, 38 - 48  keV, 48 - 65  keV energy ranges by folding the light curves of each of these ranges at the determined P$_{\mathrm{spin}}$ value. The light curves from both FPMA and FPMB were added using the \texttt{lcmath} routine prior to the folding. The pulse profiles evolve with the increase in energy  (see Fig. \ref{fig:BBpulseprofiles-all}). As also noted by  \citep{Shaw_2009}, the pulse profile has a multi-peaked structure in the lower energy ranges  (< 18 keV) while it evolves into a single-peaked structure above 18 keV. In particular, the 9 - 18 keV pulse profile has a more complex shape than the 3 - 9 keV one in that the minima of the former appear earlier between the phases 0.9-1.0 than the latter which appears between the phases 1.0-1.1. There is also another small maxima at around the phase of 0.6 apart from the biggest peak at phase $\sim 0.7$ as can be seen in panel  (2) of Fig. \ref{fig:BBpulseprofiles-all}. The pulse profiles in the ranges above 18 keV have a single-peaked structure. It is interesting to note that the minima in the 18 - 28 keV profile appears at a phase of $\sim 1.2$  (see panel  (3)), unlike the 3 - 9 keV profile. The pulse profile around the cyclotron line  (described in the next section) at 43 keV  (38 - 48 keV) is not different from those at lower energies  (18 - 28 keV and 28 - 38 keV), which is unlike some other pulsars that show abrupt changes in the shapes of the pulse profiles near the cyclotron line energy. For example, pulse profiles of XTE J1946+274  \citep{Gorban_2021} and 4U 1901+03  \citep{Beri_2021} show an evolution from a double-peaked to a single-peaked structure around the cyclotron line energy. Pulsations are detected in the 3 - 48 keV energy bands  (panel 1-5) and are not significant above 48 keV  (panel  (6)). 

\subsection{Broadband Spectral Analysis}
Due to the Ghost Rays contaminating the source region we performed the spectral analysis for all three choices of background regions  (BKG\_A, BKG\_B and BKG\_C). The results are not significantly affected by the choice of background as can be seen in Table. \ref{table:Bestfit-table-spec}. The $\chi^2$ mentioned in the following paragraph are for the choice of background region, BKG\_B.
\par 
%%%%%%%%%%%%%%%%%%%%%%%%%%%%%%%%%%%%%%%%%%%%%%%%%%%%%%%%%%%
\begin{figure*}
    \centering
	\includegraphics[scale=0.29, trim={0 3.0cm 0 1.8cm},  angle=-90]{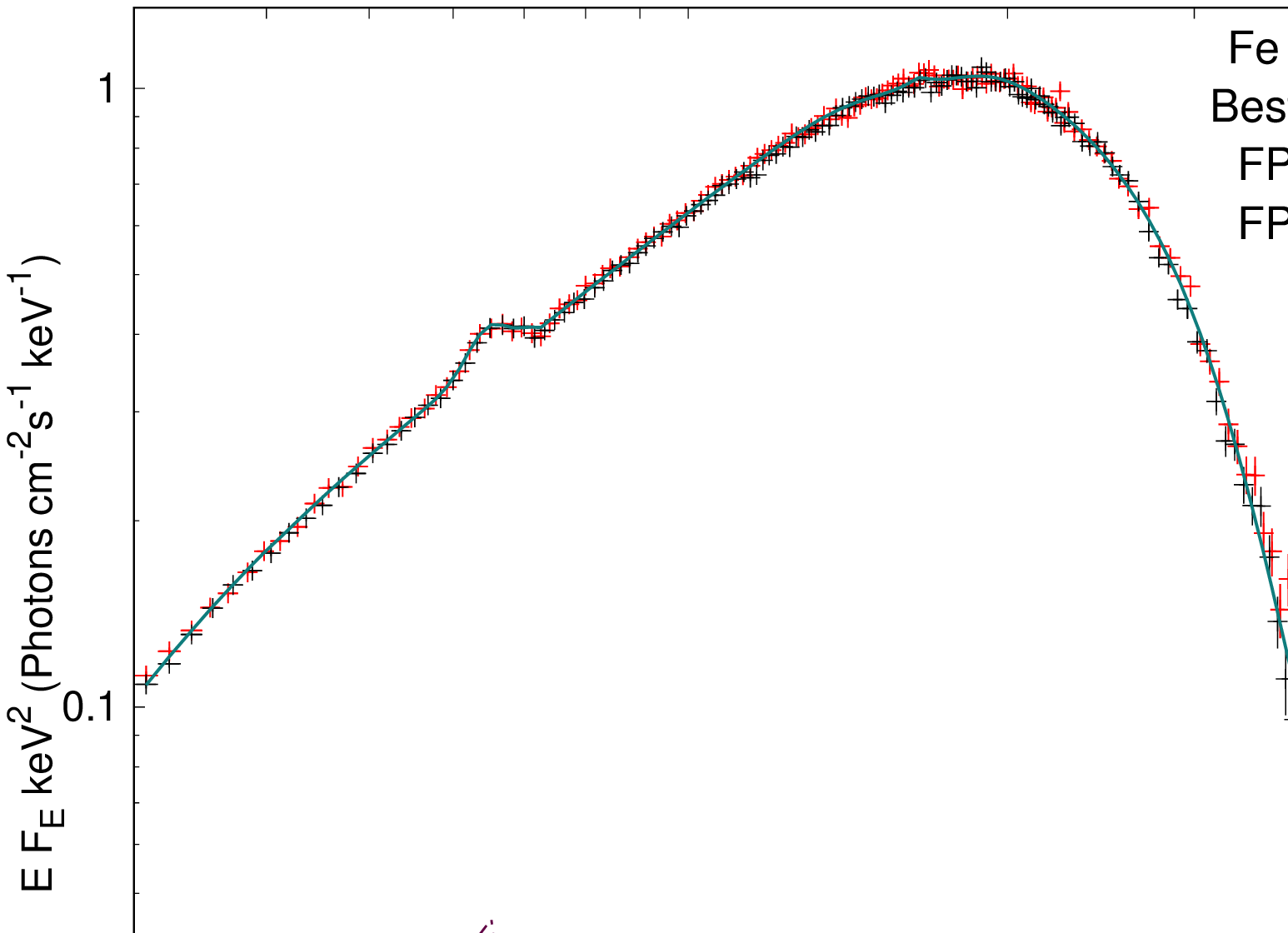}
    \caption{The unfolded spectrum with the best-fit of \texttt{HighECut} model is shown on the left panel. The right panel shows the residues. Panel  (a) represents the residues when the continuum is fit without a "43 keV \texttt{gabs}" up to 65 keV.    (b) is the best fit including a \texttt{gabs} at 43 keV and  (c) is a result of fitting the continuum up to only 27 keV, but the extended range is shown. The red and black points correspond to data from FPMA and FPMB, respectively.}
    \label{fig:unfolded-spectra-residues} 
\end{figure*}

%%%%%%%%%%%%%%%%%%%%%%%%%%%%%%%%%%%%%%%%%%%%%%%%%%%%%%%%%%%%
%%%%%%%%%%%%%%%%%%%%%%%%%%%%--------TABLE of Fitting Params-----------%%%%%%%%%%%%%%%%%%%%%%%%%%%%%%%%%%%%%%
%\begin{table*}
%\caption{Best-fitting Phase-averaged Spectral Parameters. Errors are reported at 90 \% confidence. }
\begin{table}
 \bgroup
 \def\arraystretch{1.32}
 \caption{Best-fitting Phase-averaged Spectral Parameters. Errors are reported at 90 \% confidence. }
\begin{center}

\input{table}

 \end{center}
  \label{table:Bestfit-table-spec}

 \egroup
 \begin{tablenotes}
 \item[*] $^\dagger$In units of photons cm$^{-2}$s$^{-1}$ 
 \item[*]$^*$In units of photons keV$^{-1}$cm$^{-2}$s$^{-1}$
 \item[*]$^{\ddagger}$In units of \funi [3 - 79 keV]
 \end{tablenotes}
\end{table}

%%%%%%%%%%%%%%%%%%%%%%%%%%%%%%%%%%%%%%%%%%%%%%%%%%%%%%%%%%%%%%%%%%%%%%%%%%%%%%%%%%%%%%%%%%%%%%%%%%%%%%%%%%%%

The spectrum of GRO J1750-27 resembles what is typically seen of accreting X-ray pulsars, i.e., a high energy cutoff power-law  \citep{White_1983}. In order to model the continuum, we used the model, described in \textit{XSPEC} as,\  \texttt{const $\times$ tbabs $\times$  (powerlaw $\times$ highecut )} where the \texttt{highecut} model is an exponential roll-over above a cut off energy\footnote{https://heasarc.gsfc.nasa.gov/xanadu/xspec/manual/node244.html}, \texttt{tbabs} accounts for absorbing material in the line of sight  \citep{Wilms_2000} and the \texttt{const} is used to account for the relative  normalization of the two modules. This fitting resulted in an unacceptable $\chi^2$ of 3958 for 421 d.o.f. The presence of an Fe $K_\alpha$ line at $\sim$6.4 keV is detected in the spectrum and was accounted for with a gaussian emission feature. In addition, to deal with an artificial absorption like feature around the cutoff energy  (E$_{\mathrm{cut}}\sim$16.5 keV) arising due to the slope of this function having a discontinuity, we used a smoothing gaussian absorption feature  (\texttt{gabs}) with the centroid energy tied to the cutoff and the width fixed to 1.6 keV  ($\sim$0.1 E$_{\mathrm{cut}}$) following the prescription suggested by  \citet{Coburn_2002}. The inclusion of these two features reduced the $\chi^2$ significantly to 1285 for 417 d.o.f. The residues from this fit are shown in panel  (a) of Fig. \ref{fig:unfolded-spectra-residues}. Clear residues that resemble an absorption feature around 43 keV can be seen. The inclusion of a gaussian absorption model component at this energy improved the $\chi^2$ to 500 for 414 d.o.f. The model used to find the best fit parameters was  \texttt{const $\times$ tbabs $\times$  (powerlaw $\times$ highecut + gauss) $\times$ gabs$_{\mathrm{smooth}}\times$ gabs$_{\mathrm{cyc}}$}.  We obtain an N$_\mathrm{H} \sim 2.5 \pm 0.2 \times 10^{22}$cm$^{-2}$. The residues after the best-fit are shown in panel  (b) of Fig.\ref{fig:unfolded-spectra-residues} and the best-fit parameters are given in Table \ref{table:Bestfit-table-spec}. The wavy residuals in panel  (a) below 30 keV also disappear upon including a \texttt{gabs} component at 43 keV. The presence of the wavy residuals are due to the continuum parameters without the \texttt{gabs} component being adjusted to minimize the $\chi^2$. The statistical significance as a result of this large improvement of $\chi^2$ is still high. We interpret this feature as a cyclotron absorption feature. This feature is prominent enough that it can also be seen in the raw spectrum in Fig. \ref{fig:SourceBkgSpectrum}. This feature at $43.7\pm0.9$ keV has a width of $\sim 8.0\pm0.8$  keV and a large optical depth of $\tau \sim 1.8\pm 0.2$. 
\par
The presence of the cyclotron line at 43 keV should have no effect on the continuum well below that energy. To verify this we fitted the spectrum up to 27 keV with the same continuum model excluding the cyclotron absorption feature. We then extended the same model to 65 keV and the residuals from this fit are shown in panel  (c) of Fig. \ref{fig:unfolded-spectra-residues} where the extended range has been shown. Comparing panel  (c) with panel  (b) which represents the best fit, it can be seen that the residues are similar up to 27 keV. We obtain an N$_{\mathrm{H}}$ = 2.5 $\pm 0.17$,  $\Gamma=0.78\pm0.01$, $\Gamma_{\mathrm{norm}}=3.9\pm0.1$, E$_{\mathrm{cut}}=16.8\pm0.2$ and E${_{\mathrm{fold}}}=8.7\pm0.2$ in the same units of the parameters as in Table. \ref{table:Bestfit-table-spec}. This implies that the continuum parameters are independent of the cyclotron line parameters and the dip at 43 keV is not an artifact of the modelling. Also, the presence of the cyclotron line is independent of the choice of background region among the three regions shown in Fig. \ref{fig:SourceImage}, as can be seen in Table. \ref{table:Bestfit-table-spec} where the model parameter values are given for different choices of background regions. We can also see from Fig. \ref{fig:SourceBkgSpectrum} that above 20 keV, all three backgrounds are comparable and therefore do not affect the cyclotron line parameters at 43 keV.

\section{DISCUSSION}

Despite having discovered hundreds of Neutron Star X-ray binaries, only $\sim$40 or so sources have been confirmed to exhibit cyclotron lines in their spectra  \citep{Staubert_2019}.  The cyclotron lines are indicative of the magnetic field in the region where they are formed rather than the field strength at the  surface of the NS. Using the equation $E_{\texttt{cyc}} \approx 11.6[\mathrm{keV}]\times B_{12}$, we estimate a magnetic field strength of B$\sim$3.7 $\times 10^{12}$ G for GRO J1750-27  (not accounting for the gravitational redshift) where $B_{12}$ is the magnetic field strength in the units of $10^{12}$ G. GRO J1750-27, with it's line at 43.7 keV is among sources like A0535+26 with line at 45 keV \citep{Camero-Arranz_2012}, IGR J19294+1816 with line at 43 keV \citep{Raman_2021}, GX 304-1 with line at $\sim$ 50 keV \citep{Klochkov_2012} and GX 301-2 at $\sim 50$ keV \citep{Furst_2018} that show a fundamental cyclotron line above the energy of 42 keV  .
\par
Another method that was used to estimate the magnetic field strength of the Neutron Star  (NS) makes use of the Accretion Torque model put forward by  \citet{Ghosh_1978}. The in-falling matter which has angular momentum is expected to apply a torque onto the NS at the magnetospheric radius causing the NS to either spin up or spin down. The spin change rate depends on both the period, the mass accretion rate  (Luminosity) and the magnetic field strength of the NS. Therefore, by measuring the first three quantities observationally, one can estimate a field strength for the canonical NS parameters. Using this method  \citet{Lutovinov_2019} estimate the field strength of GRO 1750-27 to be in the range of 3.5 - 4.5 $ \times 10^{12}$ G. The value that we estimate is consistent with this. However, as noted by  \citet{Kabiraj_2020}, the estimates of the magnetic field using both these methods are not always consistent. The likely contributions could be due to the difference between the canonical NS parameters that we choose and the actual scenario. The distance uncertainty would also contribute to the uncertainty in knowing the luminosity accurately.
\par
In addition to these factors, the cyclotron line parameters also show correlations with the luminosity of the source, and these correlations have been explained using several theories, of which the Shock-height model is one  \citep{Becker_2012}. This theory suggests that the height of the cyclotron line forming region varies as a function of the luminosity (L$_{\mathrm{x}}$), where one observes a positive correlation if the NS is in the Coulombic shock regime and a negative correlation if it's in the Radiation shock regime. Depending on how close to the surface or away the line forming region is, the measured magnetic field estimate may vary. The transition between these two regimes occurs at a certain luminosity, termed the critical luminosity, \Lc. Sources such as V0332+53  \citep{Lutovinov_2015} and A0535+26  \citep{Ballhausen_2017,Kong_2021} have been observed over a large range of luminosities and have shown a transition from a positive to a negative correlation. For GRO J1750-27, from this observation, we estimate a flux of 2.2 $\times 10^{-9}$\funi in the 3 - 79 keV range. The luminosity is in the range of 0.52 - 1.27 $\times 10^{38}$ erg cm$^{-2}$s$^{-1}$ if we assume the distance to be between 14  and 22 kpc \citep{Lutovinov_2019} . The source continued to brighten to almost twice it's flux until it reached the peak of the outburst as we can see in Fig. \ref{fig:outburst-lightcurve}. Even if we assume a lower distance estimate of 14 kpc, at the peak of the outburst the source would have reached a luminosity of $\sim$ 1.04 $\times 10^{38}$ erg cm$^{-2}$s$^{-1}$ which is bordering the eddington limit of a 1.4 M$\odot$ NS and puts it in the super-critical  accretion limit. Studying cyclotron line sources such as these over a wide range of luminosities, according to the present theories, we  expect to see a change in the nature of correlation (\ecyc vs \Lx) as well as changes in the pulse profiles (similar to \citealt{Wilson-Hodge_2018}) across the critical luminosity. A transition in the accretion regime may be determined if the source were to be observed over the course of future outbursts by \textit{NuSTAR} and other future hard X-ray missions.

\section*{Acknowledgements}
We thank the referee for the useful comments that improved the quality of this paper. This research made use of data obtained with \textit{NuSTAR}, a project led by Caltech, funded by NASA and managed by NASA/JPL, and it utilised the NUSTARDAS software package, jointly developed by the ASDC  (Italy) and Caltech  (USA). 

%%%%%%%%%%%%%%%%%%%%%%%%%%%%%%%%%%%%%%%%%%%%%%%%%%
\section*{Data Availability}
The data used in this paper are accessible through NASA's HEASARC website.



\bibliographystyle{mnras}
\bibliography{bibtex} % if your bibtex file is called example.bib

%\begin{thebibliography}{99}
%\bibitem[\protect \citeauthoryear{Author}{2012}]{Author2012}
%\bibitem[\protect \citeauthoryear{Others}{2013}]{Others2013}
%\end{thebibliography}

% Don't change these lines
%\bsp	% typesetting comment
\label{lastpage}
\end{document}

%% file: table.tex
\begin{tabular}{|p{2.2cm}|p{1.5cm}|p{1.5cm}|p{1.5cm}|}
 \hline
Background  &          BKG\_A                      &                           BKG\_B      &          BKG\_C                      \\
 \hline 
 \hline
C$_\mathrm{B}$   &    $1.012 \pm 0.003$       &     $1.013\pm0.003$       &     $1.013\pm0.003$       \\
N$_\mathrm{H}$ [$\times$ 10$^{22} \mathrm{cm}^{-2}$]          &    $2.44\pm0.18$            &     $2.49\pm0.18$            &     $2.56_{-0.18}^{+0.18}$            \\
 $\Gamma$&    $0.782\pm0.012$        &     $0.784\pm0.012$        &     $0.780\pm0.012$        \\
 $\Gamma_{\mathrm{norm}}$  [$\times$ 10$^{-2}]^*$     &    $3.88\pm0.11$     &     $3.90\pm0.11$     &     $3.86\pm0.11$      \\
E$_{\mathrm{cut}}$ (keV)   &    $16.45_{-0.22}^{+0.2}$            &     $16.5_{-0.2}^{+0.19}$             &     $16.43_{-0.25}^{+0.21}$           \\
E$_{\mathrm{fold}}$ kT (keV)    &    $10.29_{-0.6}^{+0.87}$            &     $10.1_{-0.54}^{+0.74}$            &     $10.27_{-0.613}^{+1}$             \\
E$_{\mathrm{Fe}}$ (keV)     &    $6.437\pm0.03$          &     $6.44\pm0.03$          &     $6.44\pm0.03$          \\
$\sigma_{\mathrm{Fe}}$  (keV)  &    $0.2584_{-0.035}^{+0.037}$        &     $0.256_{-0.035}^{+0.037}$         &     $0.254_{-0.035}^{+0.037}$        \\
Norm$_{\mathrm{Fe}}$ [$\times$ 10$^{-4}]^{\dagger}$      &    $7.57_{-0.76}^{+0.79}$  &     $7.52_{-0.75}^{+0.79}$  &     $7.42_{-0.75}^{+0.70}$  \\
E$_{\mathrm{smooth}}$ (keV)       &    16.4 (fixed)                    &     16.5 (fixed)                     &     16.4  (fixed)                   \\
$\sigma_{\mathrm{smooth}}$  (keV)    &    1.6 (fixed)                    &     1.6 (fixed)              &     1.6 (fixed)                \\
$\tau_{\mathrm{smooth}}$     &    $0.118_{-0.013}^{+0.012}$        &     $0.122\pm0.012$        &     $0.119_{-0.014}^{+0.012}$        \\
E$_{\mathrm{cyc}}$  (keV)     &    $43.87_{-0.78}^{+0.98}$           &     $43.72_{-0.75}^{+0.92}$           &     $43.69_{-0.78}^{+1.05}$           \\
$\sigma_{\mathrm{cyc}}$   (keV)    &    $8.13_{-0.65}^{+0.82}$            &     $7.97_{-0.61}^{+0.76}$            &     $8.05_{-0.66}^{+0.9}$             \\
$\tau_{\mathrm{cyc}}$     &    $1.91_{-0.18}^{+0.24}$            &     $1.84_{-0.16}^{+0.21}$            &     $1.87_{-0.18}^{+0.27}$            \\
\hline
$\chi^2$(dof) [w/o cyc]        &    1286(417)                 &     1285(417)                 &     1273(417)                  \\
$\chi^2$(dof) [w/ cyc]        &    487.2(414)                 &     500.1(414)                  &     482.9(414)                   \\

Flux [$\times 10^{-9}]^{\ddagger}$    &    $2.225\pm0.007$                      &     $2.225\pm0.007$                      &     $2.225\pm0.007$                      \\

\hline
\end{tabular}

%% file: main.bbl
\begin{thebibliography}{}
\makeatletter
\relax
\def\mn@urlcharsother{\let\do\@makeother \do\$\do\&\do\#\do\^\do\_\do\%\do\~}
\def\mn@doi{\begingroup\mn@urlcharsother \@ifnextchar [ {\mn@doi@}
  {\mn@doi@[]}}
\def\mn@doi@[#1]#2{\def\@tempa{#1}\ifx\@tempa\@empty \href
  {http://dx.doi.org/#2} {doi:#2}\else \href {http://dx.doi.org/#2} {#1}\fi
  \endgroup}
\def\mn@eprint#1#2{\mn@eprint@#1:#2::\@nil}
\def\mn@eprint@arXiv#1{\href {http://arxiv.org/abs/#1} {{\tt arXiv:#1}}}
\def\mn@eprint@dblp#1{\href {http://dblp.uni-trier.de/rec/bibtex/#1.xml}
  {dblp:#1}}
\def\mn@eprint@#1:#2:#3:#4\@nil{\def\@tempa {#1}\def\@tempb {#2}\def\@tempc
  {#3}\ifx \@tempc \@empty \let \@tempc \@tempb \let \@tempb \@tempa \fi \ifx
  \@tempb \@empty \def\@tempb {arXiv}\fi \@ifundefined
  {mn@eprint@\@tempb}{\@tempb:\@tempc}{\expandafter \expandafter \csname
  mn@eprint@\@tempb\endcsname \expandafter{\@tempc}}}

\bibitem[\protect\citeauthoryear{{Ballhausen} et~al.,}{{Ballhausen}
  et~al.}{2017}]{Ballhausen_2017}
{Ballhausen} R.,  et~al., 2017, \mn@doi [\aap] {10.1051/0004-6361/201730845},
  \href {https://ui.adsabs.harvard.edu/abs/2017A&A...608A.105B} {608, A105}

\bibitem[\protect\citeauthoryear{{Becker} et~al.,}{{Becker}
  et~al.}{2012}]{Becker_2012}
{Becker} P.~A.,  et~al., 2012, \mn@doi [\aap] {10.1051/0004-6361/201219065},
  \href {https://ui.adsabs.harvard.edu/abs/2012A&A...544A.123B} {544, A123}

\bibitem[\protect\citeauthoryear{{Beri}, {Girdhar}, {Iyer}  \& {Maitra}}{{Beri}
  et~al.}{2021}]{Beri_2021}
{Beri} A.,  {Girdhar} T.,  {Iyer} N.~K.,   {Maitra} C.,  2021, \mn@doi [\mnras]
  {10.1093/mnras/staa3345}, \href
  {https://ui.adsabs.harvard.edu/abs/2021MNRAS.500.1350B} {500, 1350}

\bibitem[\protect\citeauthoryear{{Camero-Arranz} et~al.,}{{Camero-Arranz}
  et~al.}{2012}]{Camero-Arranz_2012}
{Camero-Arranz} A.,  et~al., 2012, \mn@doi [\apj] {10.1088/0004-637X/754/1/20},
  \href {https://ui.adsabs.harvard.edu/abs/2012ApJ...754...20C} {754, 20}

\bibitem[\protect\citeauthoryear{{Coburn}, {Heindl}, {Rothschild}, {Gruber},
  {Kreykenbohm}, {Wilms}, {Kretschmar}  \& {Staubert}}{{Coburn}
  et~al.}{2002}]{Coburn_2002}
{Coburn} W.,  {Heindl} W.~A.,  {Rothschild} R.~E.,  {Gruber} D.~E.,
  {Kreykenbohm} I.,  {Wilms} J.,  {Kretschmar} P.,   {Staubert} R.,  2002,
  \mn@doi [\apj] {10.1086/343033}, \href
  {https://ui.adsabs.harvard.edu/abs/2002ApJ...580..394C} {580, 394}

\bibitem[\protect\citeauthoryear{{Corbet}}{{Corbet}}{1984}]{corbet_1984}
{Corbet} R.~H.~D.,  1984, \aap, \href
  {https://ui.adsabs.harvard.edu/abs/1984A&A...141...91C} {141, 91}

\bibitem[\protect\citeauthoryear{{Finger} \& {Wilson-Hodge}}{{Finger} \&
  {Wilson-Hodge}}{2014}]{Finger_2014}
{Finger} M.~H.,  {Wilson-Hodge} C.~A.,  2014, The Astronomer's Telegram, \href
  {https://ui.adsabs.harvard.edu/abs/2014ATel.6839....1F} {6839, 1}

\bibitem[\protect\citeauthoryear{{F{\"u}rst} et~al.,}{{F{\"u}rst}
  et~al.}{2018}]{Furst_2018}
{F{\"u}rst} F.,  et~al., 2018, \mn@doi [\aap] {10.1051/0004-6361/201732132},
  \href {https://ui.adsabs.harvard.edu/abs/2018A&A...620A.153F} {620, A153}

\bibitem[\protect\citeauthoryear{{Ghosh} \& {Lamb}}{{Ghosh} \&
  {Lamb}}{1978}]{Ghosh_1978}
{Ghosh} P.,  {Lamb} F.~K.,  1978, \mn@doi [\apjl] {10.1086/182734}, \href
  {https://ui.adsabs.harvard.edu/abs/1978ApJ...223L..83G} {223, L83}

\bibitem[\protect\citeauthoryear{{Gorban}, {Molkov}, {Tsygankov}  \&
  {Lutovinov}}{{Gorban} et~al.}{2021}]{Gorban_2021}
{Gorban} A.~S.,  {Molkov} S.~V.,  {Tsygankov} S.~S.,   {Lutovinov} A.~A.,
  2021, \mn@doi [Astronomy Letters] {10.1134/S1063773721060049}, \href
  {https://ui.adsabs.harvard.edu/abs/2021AstL...47..390G} {47, 390}

\bibitem[\protect\citeauthoryear{Harrison et~al.,}{Harrison
  et~al.}{2013}]{Harrison_2013}
Harrison F.~A.,  et~al., 2013, \mn@doi [The Astrophysical Journal]
  {10.1088/0004-637x/770/2/103}, 770, 103

\bibitem[\protect\citeauthoryear{{Kaastra} \& {Bleeker}}{{Kaastra} \&
  {Bleeker}}{2016}]{Kaastra_2016}
{Kaastra} J.~S.,  {Bleeker} J.~A.~M.,  2016, \mn@doi [\aap]
  {10.1051/0004-6361/201527395}, \href
  {https://ui.adsabs.harvard.edu/abs/2016A&A...587A.151K} {587, A151}

\bibitem[\protect\citeauthoryear{{Kabiraj} \& {Paul}}{{Kabiraj} \&
  {Paul}}{2020}]{Kabiraj_2020}
{Kabiraj} S.,  {Paul} B.,  2020, \mn@doi [\mnras] {10.1093/mnras/staa2079},
  \href {https://ui.adsabs.harvard.edu/abs/2020MNRAS.497.1059K} {497, 1059}

\bibitem[\protect\citeauthoryear{{Klochkov} et~al.,}{{Klochkov}
  et~al.}{2012}]{Klochkov_2012}
{Klochkov} D.,  et~al., 2012, \mn@doi [\aap] {10.1051/0004-6361/201219385},
  \href {https://ui.adsabs.harvard.edu/abs/2012A&A...542L..28K} {542, L28}

\bibitem[\protect\citeauthoryear{{Koh}, {Chakrabarty}, {Prince}, {Vaughan},
  {Zhang}, {Scott}, {Finger}  \& {Wilson}}{{Koh} et~al.}{1995}]{Koh_1995}
{Koh} T.,  {Chakrabarty} D.,  {Prince} T.~A.,  {Vaughan} B.,  {Zhang} S.~N.,
  {Scott} M.,  {Finger} M.~H.,   {Wilson} R.~B.,  1995, \iaucirc, \href
  {https://ui.adsabs.harvard.edu/abs/1995IAUC.6222....1K} {6222, 1}

\bibitem[\protect\citeauthoryear{{Kong} et~al.,}{{Kong}
  et~al.}{2021}]{Kong_2021}
{Kong} L.~D.,  et~al., 2021, arXiv e-prints, \href
  {https://ui.adsabs.harvard.edu/abs/2021arXiv210802485K} {p. arXiv:2108.02485}

\bibitem[\protect\citeauthoryear{{Lutovinov}, {Tsygankov}, {Suleimanov},
  {Mushtukov}, {Doroshenko}, {Nagirner}  \& {Poutanen}}{{Lutovinov}
  et~al.}{2015}]{Lutovinov_2015}
{Lutovinov} A.~A.,  {Tsygankov} S.~S.,  {Suleimanov} V.~F.,  {Mushtukov} A.~A.,
   {Doroshenko} V.,  {Nagirner} D.~I.,   {Poutanen} J.,  2015, \mn@doi [\mnras]
  {10.1093/mnras/stv125}, \href
  {https://ui.adsabs.harvard.edu/abs/2015MNRAS.448.2175L} {448, 2175}

\bibitem[\protect\citeauthoryear{{Lutovinov}, {Tsygankov}, {Karasev}, {Molkov}
  \& {Doroshenko}}{{Lutovinov} et~al.}{2019}]{Lutovinov_2019}
{Lutovinov} A.~A.,  {Tsygankov} S.~S.,  {Karasev} D.~I.,  {Molkov} S.~V.,
  {Doroshenko} V.,  2019, \mn@doi [\mnras] {10.1093/mnras/stz437}, \href
  {https://ui.adsabs.harvard.edu/abs/2019MNRAS.485..770L} {485, 770}

\bibitem[\protect\citeauthoryear{{Madsen}, {Christensen}, {Craig}, {Forster},
  {Grefenstette}, {Harrison}, {Miyasaka}  \& {Rana}}{{Madsen}
  et~al.}{2017}]{Madsen_2017}
{Madsen} K.~K.,  {Christensen} F.~E.,  {Craig} W.~W.,  {Forster} K.~W.,
  {Grefenstette} B.~W.,  {Harrison} F.~A.,  {Miyasaka} H.,   {Rana} V.,  2017,
  \mn@doi [Journal of Astronomical Telescopes, Instruments, and Systems]
  {10.1117/1.JATIS.3.4.044003}, \href
  {https://ui.adsabs.harvard.edu/abs/2017JATIS...3d4003M} {3, 044003}

\bibitem[\protect\citeauthoryear{{Malacaria}, {Jenke}  \&
  {Wilson-Hodge}}{{Malacaria} et~al.}{2021}]{Malacaria_2021}
{Malacaria} C.,  {Jenke} P.,   {Wilson-Hodge} C.,  2021, The Astronomer's
  Telegram, \href {https://ui.adsabs.harvard.edu/abs/2021ATel14930....1M}
  {14930, 1}

\bibitem[\protect\citeauthoryear{{Raman}, {Varun,}, {Paul}  \&
  {Bhattacharya}}{{Raman} et~al.}{2021}]{Raman_2021}
{Raman} G.,  {Varun,} {Paul} B.,   {Bhattacharya} D.,  2021, \mn@doi [\mnras]
  {10.1093/mnras/stab2835}, \href
  {https://ui.adsabs.harvard.edu/abs/2021MNRAS.508.5578R} {508, 5578}

\bibitem[\protect\citeauthoryear{{Reig}}{{Reig}}{2011}]{Reig_2011}
{Reig} P.,  2011, \mn@doi [\apss] {10.1007/s10509-010-0575-8}, \href
  {https://ui.adsabs.harvard.edu/abs/2011Ap&SS.332....1R} {332, 1}

\bibitem[\protect\citeauthoryear{{Rouco Escorial}, {Wijnands}, {Ootes},
  {Degenaar}, {Snelders}, {Kaper}, {Cackett}  \& {Homan}}{{Rouco Escorial}
  et~al.}{2019}]{Escorial_2019}
{Rouco Escorial} A.,  {Wijnands} R.,  {Ootes} L.~S.,  {Degenaar} N.,
  {Snelders} M.,  {Kaper} L.,  {Cackett} E.~M.,   {Homan} J.,  2019, \mn@doi
  [\aap] {10.1051/0004-6361/201834327}, \href
  {https://ui.adsabs.harvard.edu/abs/2019A&A...630A.105R} {630, A105}

\bibitem[\protect\citeauthoryear{{Scott}, {Finger}, {Wilson}, {Koh}, {Prince},
  {Vaughan}  \& {Chakrabarty}}{{Scott} et~al.}{1997}]{Scott_1997}
{Scott} D.~M.,  {Finger} M.~H.,  {Wilson} R.~B.,  {Koh} D.~T.,  {Prince} T.~A.,
   {Vaughan} B.~A.,   {Chakrabarty} D.,  1997, \mn@doi [\apj] {10.1086/304740},
  \href {https://ui.adsabs.harvard.edu/abs/1997ApJ...488..831S} {488, 831}

\bibitem[\protect\citeauthoryear{{Shaw}, {Hill}, {Kuulkers}, {Brandt},
  {Chenevez}  \& {Kretschmar}}{{Shaw} et~al.}{2009}]{Shaw_2009}
{Shaw} S.~E.,  {Hill} A.~B.,  {Kuulkers} E.,  {Brandt} S.,  {Chenevez} J.,
  {Kretschmar} P.,  2009, \mn@doi [\mnras] {10.1111/j.1365-2966.2008.14212.x},
  \href {https://ui.adsabs.harvard.edu/abs/2009MNRAS.393..419S} {393, 419}

\bibitem[\protect\citeauthoryear{{Staubert} et~al.,}{{Staubert}
  et~al.}{2019}]{Staubert_2019}
{Staubert} R.,  et~al., 2019, \mn@doi [\aap] {10.1051/0004-6361/201834479},
  \href {https://ui.adsabs.harvard.edu/abs/2019A&A...622A..61S} {622, A61}

\bibitem[\protect\citeauthoryear{{White}, {Swank}  \& {Holt}}{{White}
  et~al.}{1983}]{White_1983}
{White} N.~E.,  {Swank} J.~H.,   {Holt} S.~S.,  1983, \mn@doi [\apj]
  {10.1086/161162}, \href
  {https://ui.adsabs.harvard.edu/abs/1983ApJ...270..711W} {270, 711}

\bibitem[\protect\citeauthoryear{{Wilms}, {Allen}  \& {McCray}}{{Wilms}
  et~al.}{2000}]{Wilms_2000}
{Wilms} J.,  {Allen} A.,   {McCray} R.,  2000, \mn@doi [\apj] {10.1086/317016},
  \href {https://ui.adsabs.harvard.edu/abs/2000ApJ...542..914W} {542, 914}

\bibitem[\protect\citeauthoryear{{Wilson-Hodge} et~al.,}{{Wilson-Hodge}
  et~al.}{2018}]{Wilson-Hodge_2018}
{Wilson-Hodge} C.~A.,  et~al., 2018, \mn@doi [\apj] {10.3847/1538-4357/aace60},
  \href {https://ui.adsabs.harvard.edu/abs/2018ApJ...863....9W} {863, 9}

\makeatother
\end{thebibliography}
